\documentclass[final,3p,times]{elsarticle}

\usepackage{amssymb}
\usepackage{amsmath}
\usepackage{bm}

\journal{Journal of Subatomic Particles and Cosmology}

\begin{document}

\begin{frontmatter}



\title{Gaussian vs. Real Wavefunction of Nuclear Clusters and Hypernuclei}

\author[aaa,bbb]{Jiaxing Zhao}
\author[ddd]{Joerg Aichelin}
\author[ccc,aaa,bbb]{Elena Bratkovskaya}

\affiliation[aaa]{organization={Helmholtz Research Academy Hessen for FAIR (HFHF)},
             addressline={GSI Helmholtz Center for Heavy Ion Physics. Campus Frankfurt},
             city={Frankfurt am Main},
             postcode={60438},
             country={Germany}}

 \affiliation[bbb]{organization={Institute for Theoretical Physics, Johann Wolfgang Goethe University},
             addressline={Max-von-Laue-Str. 1},
             city={Frankfurt am Main},
             postcode={60438},
             country={Germany}}
                       
 \affiliation[ccc]{organization={GSI Helmholtzzentrum fur Schwerionenforschung GmbH},
             addressline={Planckstr. 1},
             city={Darmstadt},
             postcode={64291},
             country={Germany}}  
                                             
 \affiliation[ddd]{organization={SUBATECH, Nantes University, IMT Atlantique, IN2P3/CNRS},
             addressline={4 rue Alfred Kastler},
             city={Nantes},
             postcode={44307},
             country={France}}        
                                       
\begin{abstract}
We compare realistic $N$-body wave functions obtained from solutions of the Schrödinger equation with Gaussian ansätze constrained to the same rms radius. The microscopic wave functions exhibit significantly broader spatial distributions, revealing pronounced non-Gaussian structures. In addition, we investigate possible production channels for $A=4$ clusters using a phenomenological two-body interaction. This study provides a potential mechanism that may help alleviate the underestimation of $A=4$ cluster yields in theoretical models compared to experimental data.
\end{abstract}




\end{frontmatter}



Light nuclei and hypernuclei produced in relativistic heavy-ion collisions provide sensitive probes of baryon clustering, chemical freeze-out dynamics, and strangeness production. Measurements by the STAR and ALICE collaborations have established a comprehensive experimental database spanning a wide range of collision energies, while future facilities such as FAIR, J-PARC, NICA, and HIAF will further explore the high-baryon-density regime and multi-strange sector.

The production of nuclear clusters and hypernuclei in heavy-ion collisions is a complex process, involving the formation of few-body bound states within a rapidly evolving many-body environment. To date, their formation in semiclassical transport approaches is described only through phenomenological mechanisms. Three main classes of approaches have been developed: (i) {\it Coalescence}; (ii) {\it Collision-induced formation}; (iii){\it Potential-driven formation}. In all cases, the production probability is closely related to the underlying cluster wave function and binding properties.

Significant progress in ab initio nuclear many-body methods has enabled accurate calculations of bound-state properties of nuclei and hypernuclei~\cite{Leidemann:2012hr,Ekstrom:2022yea}. However, these studies are primarily focused on spectra and binding energies, whereas coalescence-based production models require the corresponding Wigner phase-space distributions. Extending realistic Wigner-function descriptions to heavier clusters and hypernuclei therefore remains an important challenge for a quantitative understanding of cluster production in $pp$ and heavy-ion collisions.

In this proceeding, we present two aspects of our study. First, we compare previously obtained microscopic wave functions with commonly used Gaussian ansätze, providing insight into the internal structure of nuclear clusters and hypernuclei. Second, we explore mass criterion to identify and estimate possible production channels. Together, these studies aim to improve our understanding of nuclear cluster and hypernuclei production from small to large collision systems, with particular emphasis on nuclei with $A>3$.

\begin{figure}[t]
\centering
\includegraphics[width=0.9\textwidth]{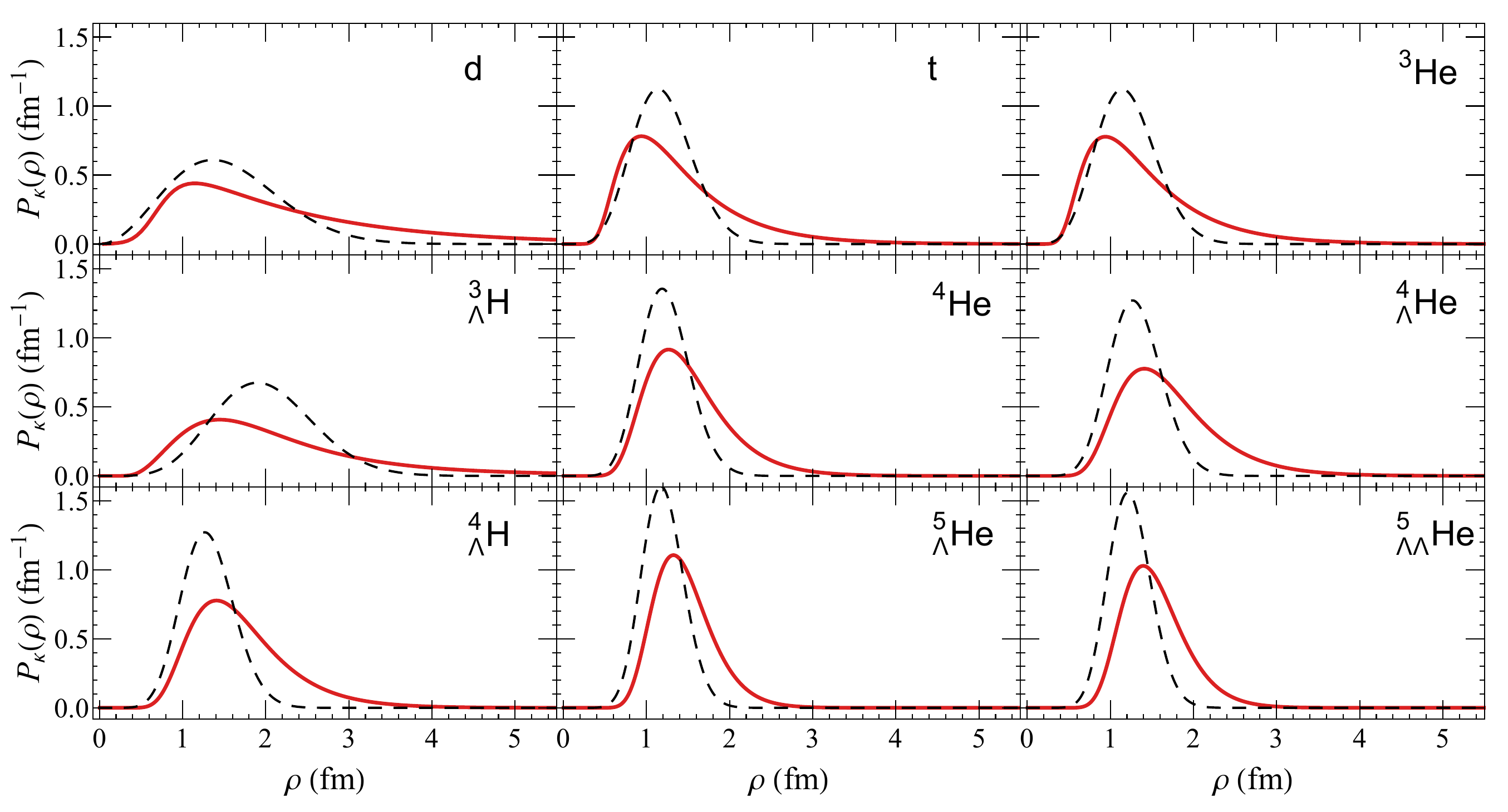}
\caption{The radial probability distributions $P_\kappa(\rho)\equiv |R_\kappa(\rho)|^2\rho^{3N-4}$ of $S$-wave $d$, $t$, $^3\rm He$, $^3_\Lambda \rm H$, $^4\rm He$, $^4_\Lambda \rm He$, $^4_\Lambda \rm H$, $^5_\Lambda \rm He$, and $^5_{\Lambda\Lambda} \rm He$ are shown as red solid lines. The corresponding Gaussian wave-function probabilities are shown as black dashed lines for comparison.}
\label{fig.1}
\end{figure}

In a previous study~\cite{Zhao:2025glf}, we solved the $N$-body Schrödinger equation within the hyperspherical harmonics formalism and computed the corresponding Wigner phase-space distributions of nuclear clusters and hypernuclei. The Hamiltonian of the $N$-body system is given by
\begin{eqnarray}
H=\sum_{i=1}^N{{\bf p}^2_i \over 2m_i}+\sum_{i<j} V_{ij}.
\end{eqnarray}
We assume that the interaction potential can be written as $V=\sum_{i<j}V_{ij}$, i.e., as a sum of two-body interactions, while genuine three-body forces are neglected. In the present calculation, we employ the \textit{Argonne}-18 potential for nucleon–nucleon interactions and the Usmani potential for hyperon–nucleon ($Y$–$N$) interactions.
To facilitate the solution of the many-body problem, we introduce Jacobi coordinates, which transform the individual particle coordinates into the center-of-mass (CoM) coordinate ${\bf R}$ and a set of relative coordinates ${\bm \chi}$~\cite{Barnea:1999be,Marcucci:2019hml,Zhao:2020nwy},
\begin{eqnarray}
{\bf R}&=& {1\over M}\sum_{i=1}^Nm_i{\bf r}_i, \nonumber\\
{\bm \chi}_{N-j}&=&\sqrt{M_j m_{j+1}\over M_{j+1}\mu }\left( {\bf r}_{j+1}-{1\over M_j}\sum_{i=1}^j m_i {\bf r}_i\right).
\end{eqnarray}
where $M_j=\sum_{i=1}^j m_i$ and $j=1,...N-1$. $\mu\equiv M$ is a parameter with the mass dimension. Its value does not affect the final results.
Further we transformed the relative coordinates ${\bm \chi}_1$,...,${\bm \chi}_{N-1}$ into one hyperradius, 
$\rho\equiv \sqrt{{\bm \chi}_1^2+...+{\bm \chi}_{N-1}^2}$
and $3N-4$ hyperangles $\Omega=\{\alpha_{N-1},...,\alpha_2, \theta_1,\phi_1,...,\theta_{N-1},\phi_{N-1}\}$.
Then the relative wavefunction satisfies the equation, 
\begin{eqnarray}
\left [ {1\over 2\mu}\left( -{1 \over \rho^{3N-4}}{d\over d\rho}\rho^{3N-4}{d \over d\rho}  + {\hat {\bf K}_N^2\over \rho^2}\right) + V(\rho, \Omega) \right ]\Phi= E_r \Phi,
\label{eq.relamom}
\end{eqnarray}
where $\hat {\bf K}_N$ is the hyperangular momentum operator. Its eigenstates is hyperspherical harmonic (HH) functions ${\mathcal Y}_\kappa(\Omega)$, which form a complete orthonormal basis in the hyperangular space. We expanded the wave function in this basis, $\Phi(\rho, \Omega)=\sum_\kappa R_{\kappa}(\rho){\mathcal Y}_\kappa(\Omega)$. $\kappa$ stands for all the quantum numbers, $\{n_i,l_i,K,L,M\}$.
The radial probability of each component is defined as,
$P_\kappa(\rho)\equiv |R_\kappa(\rho)|^2\rho^{3N-4}$, as shown in Fig.~\ref{fig.1}.
The normalization condition is fulfilled, $\int \sum_\kappa P_\kappa(\rho)d\rho=1$.
The root-mean-squared (rms) radius of the $N$-body cluster is defined as $
r_{\rm rms}^2\equiv \langle \rho^2\rangle=\int \sum_\kappa |R_\kappa(\rho)|^2\rho^{3N-2} d\rho$.

If the interaction between two nucleons is a 3-D isotropic harmonic oscillator (HO) potential, 
then the $N$-body wavefunction has analytical solution. It is written as follows:
\begin{eqnarray}
\Phi(\rho, \Omega)=\left( 2n_\rho! \over \sigma^{3(N-1)+2K}\Gamma[n_\rho+K+3(N-1)/2] \right)^{1/2} \rho^K L_{n_\rho}^{K+{3N-5\over 2}}\Big({\rho^2\over \sigma^2}\Big) e^{-\rho^2/(2\sigma^2)}{\mathcal Y}_\kappa(\Omega),
\end{eqnarray}
where $L_n^\alpha$ generalized (associated) Laguerre polynomial.
The ground state gives, 
\begin{eqnarray}
\Phi_0(\rho, \Omega)&=&\left({1\over \pi \sigma^2}\right)^{3(N-1)\over 4}\exp\left({-{\rho^2\over 2\sigma^2}}\right), \nonumber\\
\mathcal Y_{0}(\Omega)&=&\left(\Gamma[3(N-1)/2]\over 2 \pi^{3(N-1)/2}\right)^{1/2}.
\end{eqnarray}
In Fig.~\ref{fig.1}, we compare the dominant (first) component of the real wave function with a Gaussian wave function having the same rms radius. For an $N$-body Gaussian wave function, the rms radius is related to the width parameter $\sigma$ by $r_{\rm rms}^2={3(N-1)\over 2}\sigma^2$. It is evident that the realistic wave functions exhibit significantly broader spatial structures than the corresponding Gaussian ansatz. This deviation may have important consequences for the production of nuclear clusters and hypernuclei in small systems, such as $pp$, $pA$, and peripheral heavy-ion collisions~\cite{Leung:2025jwe,Mahlein:2023fmx}, where the sensitivity to the detailed structure of the wave function is expected to be enhanced.

We now discuss possible production channels for heavy clusters and hypernuclei within our framework. In particular, whether ${}^3_\Lambda \mathrm{H}$ can be interpreted as a bound state of $d$ and $\Lambda$ has been widely debated in the literature~\cite{Kamada:1997rv,Gal:2016boi}. 
In our previous calculations, the combined mass of the $d$ and $\Lambda$ system is found to be smaller than that of ${}^3_\Lambda \mathrm{H}$, which disfavor a bound $d-\Lambda$ molecular configuration within the present setup. However, this conclusion should be viewed with caution. In particular, the current treatment does not explicitly include spin-dependent forces or more sophisticated three-body correlations, which may modify the effective interaction and potentially allow for the existence of a weakly bound or near-threshold configuration.

\begin{figure}[t]
\centering
\includegraphics[width=0.4\textwidth]{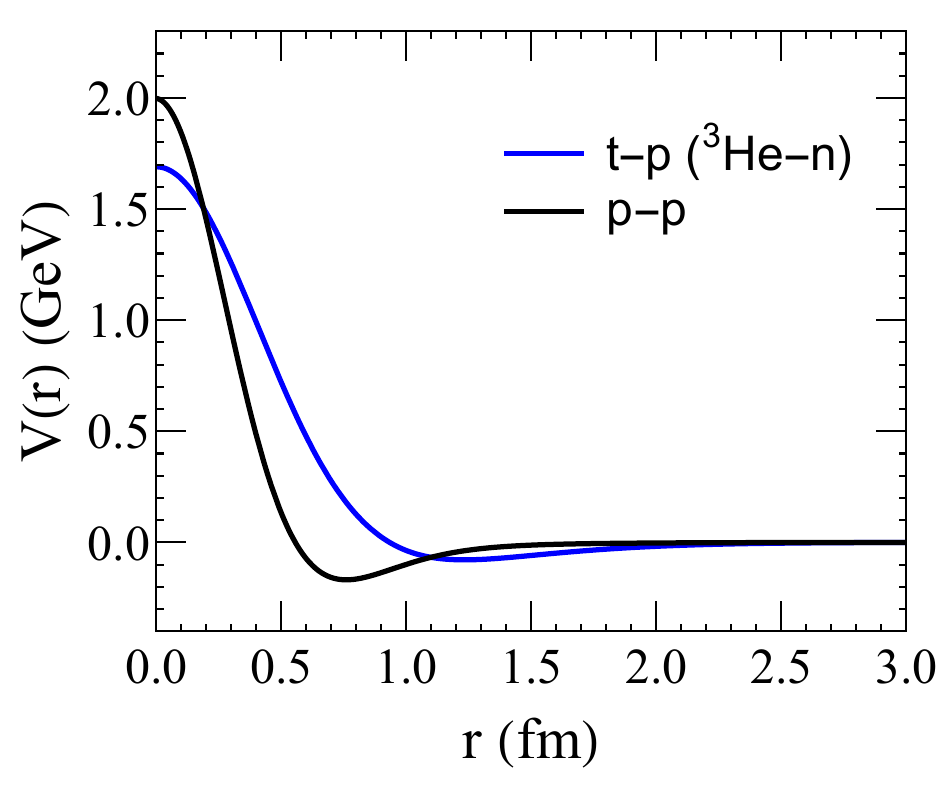}\includegraphics[width=0.4\textwidth]{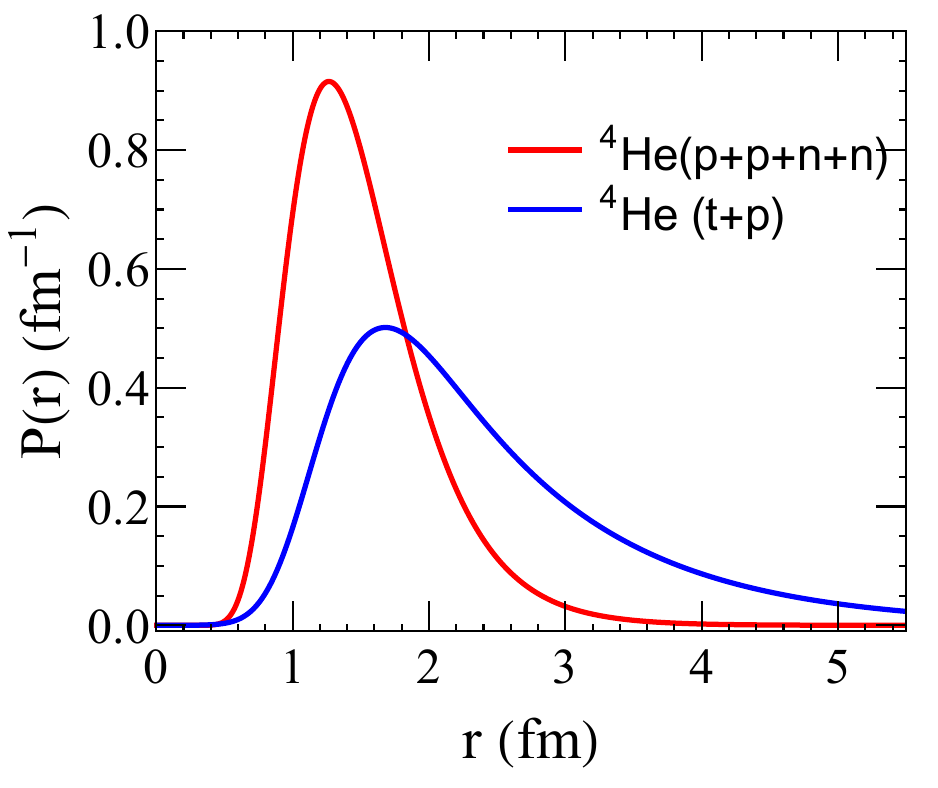}
\caption{The interaction potential of $p-p$ from \textit{Argonne}-18 and $t-p$ by fitting the mass of the $^4\rm He$ (left panel). The radial distribution $P(r)$ of $S$-wave $^4\rm He$ from 4body system $p-p-n-n$ and 2body system $t-p$ (right panel).}
\label{fig.2}
\end{figure}
We next consider possible channels for $A=4$ clusters. For $^4\mathrm{He}$, relevant two-body channels include $t+p$ and $^3\mathrm{He}+n$. Both satisfy a simple mass criterion, $m_t+m_p \gtrsim m_{^4\mathrm{He}}$ and $m_{^3\mathrm{He}}+m_n \gtrsim m_{^4\mathrm{He}}$. Since the effective interactions in these channels are not well constrained, we employ a phenomenological two-range Gaussian potential~\cite{Morita:2014kza},
\begin{equation}
V(r)=v_r\exp\left (-{r^2\over \mu_r^2}\right) + v_a\exp\left (-{r^2\over \mu_a^2}\right),
\end{equation}
representing short-range repulsion (first term) and intermediate-range attraction (second term), respectively.
The parameters are tuned by solving the two-body Schrödinger equation to reproduce the empirical $^4\mathrm{He}$ mass, yielding $v_r = 2~\mathrm{GeV}$, $v_a = -0.31~\mathrm{GeV}$, $\mu_r = 3~\mathrm{GeV}^{-1}$, and $\mu_a = 6~\mathrm{GeV}^{-1}$. The resulting potential and wave function are shown in Fig.~\ref{fig.2}. 
Due to its smaller binding energy, the $t-p$ bound state exhibits a broader spatial distribution.
Within this approach, the $t-p$ and $^3\mathrm{He}-n$ systems are treated equivalently due to their similar masses.
This provides a phenomenological estimate of possible production channels, which may help address the underestimation of $A=4$ cluster yields in theoretical models compared to experimental data~\cite{ALICE:2023qyl,STAR:2023uxk}.

In summary, we compare realistic $N$-body wave functions obtained from the $N$-body Schrödinger equation with Gaussian ansätze constrained to the same root-mean-square radius. We find that the realistic wave functions exhibit significantly broader spatial distributions than the corresponding Gaussian parametrization, revealing pronounced non-Gaussian structures in the internal dynamics of nuclear clusters and hypernuclei. Such deviations indicate that commonly used Gaussian assumptions may not fully capture the spatial correlations encoded in few-body bound states. These differences may have important consequences for the production of nuclear clusters and hypernuclei in small collision systems, such as $pp$, $pA$, and peripheral heavy-ion collisions, where the coalescence process is particularly sensitive to the detailed phase-space structure of the underlying wave functions. In this context, non-Gaussian tails can lead to nontrivial modifications of cluster formation probabilities and may affect the interpretation of experimental yields. 

In addition, we investigate possible $A=4$ production channels using a phenomenological two-body interaction model. These channels may help alleviate the long-standing underestimation of $A=4$ cluster production in theoretical models compared to experimental data. Overall, our results highlight the importance of realistic wave functions and additional production channels for a quantitative understanding of light nuclei and hypernuclei formation.

\bibliographystyle{elsarticle-num}
\bibliography{sqm2026_jxzhao}



\end{document}